\documentclass[prl,twocolumn,floatfix,superscriptaddress]{revtex4}
\usepackage{amsmath}
\usepackage{graphicx}
\usepackage{natbib}
\usepackage{subfigure}
\usepackage[colorlinks=true]{hyperref}

\begin{document}

\title{Pseudogap temperature as a Widom line in doped Mott insulators}
\author{G. Sordi}
\affiliation{Theory Group, Institut Laue Langevin, 6 rue Jules Horowitz, 38042 Grenoble Cedex, France}
\author{P. S\'emon}
\affiliation{D\'epartement de physique and Regroupement qu\'eb\'equois sur les mat\'eriaux de pointe, Universit\'e de Sherbrooke, Sherbrooke, Qu\'ebec, Canada J1K 2R1}
\author{K. Haule}
\affiliation{Department of Physics \& Astronomy, Rutgers University, Piscataway, NJ 08854-8019, USA}
\author{A.-M. S. Tremblay}
\affiliation{D\'epartement de physique and Regroupement qu\'eb\'equois sur les mat\'eriaux de pointe, Universit\'e de Sherbrooke, Sherbrooke, Qu\'ebec, Canada J1K 2R1}
\affiliation{Canadian Institute for Advanced Research, Toronto, Ontario, Canada, M5G 1Z8}

\date{\today}

\begin{abstract}
The pseudogap refers to an enigmatic state of matter with unusual physical properties found below a characteristic temperature $T^*$  in hole-doped high-temperature superconductors. 
Determining $T^*$ is critical for understanding this state.
Here we study the simplest model of correlated electron systems, the Hubbard model, with cluster dynamical mean-field theory to find out whether the pseudogap can occur solely because of strong coupling physics and short nonlocal correlations. 
We find that the pseudogap characteristic temperature $T^*$ is a sharp crossover between different dynamical regimes along a line of thermodynamic anomalies that appears above a first-order phase transition, the Widom line. 
The Widom line emanating from the critical endpoint of a first-order transition is thus the organizing principle for the pseudogap phase diagram of the cuprates. 
No additional broken symmetry is necessary to explain the phenomenon. 
Broken symmetry states appear in the pseudogap and not the other way around.
\end{abstract}
\maketitle

{\noindent\bf Introduction}\\
The phase diagram of hole-doped high-temperature superconductors remains puzzling.
A state of matter with unusual physical properties, dubbed ``the pseudogap'', is found below a characteristic temperature $T^*$ in a doped
Mott insulator.
Since the superconducting state is born out of the pseudogap over much of the phase diagram, the nature of the pseudogap is a fundamental issue in the field and it is under intense theoretical~\cite{normanADV} and experimental~\cite{Daou:2010,Lawler:2010,He:2011} scrutiny.

A pseudogap can occur because of disorder-broadened long-range ordered phases of Ising type~\cite{Fradkin:2010} or because of fluctuating precursors to a long-range ordered phase that would appear only at $T=0$ because of the Mermin-Wagner theorem~\cite{Vilk:1997}.
The phenomenology of the pseudogap in the less strongly-coupled electron-doped cuprates~\cite{st,Weber:2010} differs from that of hole-doped cuprates and is consistent with fluctuating precursors~\cite{Motoyama:2007}.
The above two generic mechanisms have in common that they can occur even when interactions between electrons are not strong enough to lead to a Mott insulator at half-filling.
Various broken symmetry states, such as spin-charge density wave, have been linked to the pseudogap in some cuprate families~\cite{Motoyama:2007,Daou:2010,Lawler:2010,He:2011} but not in all of them.
Yet the pseudogap is a generic feature of {\it all} hole-doped cuprates.
It is also possible that broken symmetries are different phenomena, and in particular only a consequence, and not the origin, of the pseudogap, as suggested recently~\cite{Parker:2010}.

On the other hand, the Mott phenomenon, a blocking of charge transport because of strong electronic repulsion, is ubiquitous in hole-doped cuprates~\cite{lee} and it is appropriate to ask whether there is a {\it third} generic mechanism for the pseudogap that is associated purely with Mott physics in two dimensions~\cite{st}.
This mechanism appears only if the Coulomb repulsion is strong enough~\cite{st} to turn the system into a Mott insulator at zero doping, as observed in hole-doped cuprates.
No broken translational or rotational symmetry is needed, although broken symmetry may also occur in certain cases~\cite{Fradkin:2010}.
In previous work~\cite{sht,sht2}, we identified a first-order transition at finite doping between two different metals. 
Here we show that the characteristic temperature $T^*$ is an unexpected example of a phenomenon observed in fluids~\cite{water1}, namely a sharp crossover between different {\it dynamical} regimes along a line of {\it thermodynamic} anomalies that appears above that first-order phase transition, the Widom line~\cite{water1}.

{\noindent\bf Results}\\
{\it Widom line in doped Mott insulator.}
To investigate the formation of the pseudogap upon doping a Mott insulator, we study the competition between nearest neighbors hopping $t$ and screened Coulomb repulsion $U$ embodied in the two-dimensional Hubbard model.
For the cuprates, the clear momentum dependence of the self-energy observed in photoemission (ARPES) forces one to use cluster extensions~\cite{kotliarRMP,maier} of dynamical mean-field theory~\cite{rmp}. It is known that four sites~\cite{kyung} suffices to reproduce qualitatively the ARPES spectrum. Larger cluster sizes~\cite{michelCFR} will improve momentum resolutions and change quantitative details but should not remove first-order transitions, as has been verified at half-filling~\cite{phk}. 
As in any cluster mean-field theory (dynamical or not), we can study the normal-state phase of the model by suppressing long-range magnetic, superconducting or other types of order while retaining the short-range correlations.
Previous works based on this theoretical framework have shown that a pseudogap appears close to a Mott insulator, as a result of short-range correlations~\cite{st,kyung,Macridin:2006,hauleDOPING}.
But it is only with recent theoretical and computational advances~\cite{millisRMP} that detailed results in the low temperature region of the phase diagram can be obtained.

\begin{figure}
\centering{
\includegraphics[width=0.99\linewidth,clip=]{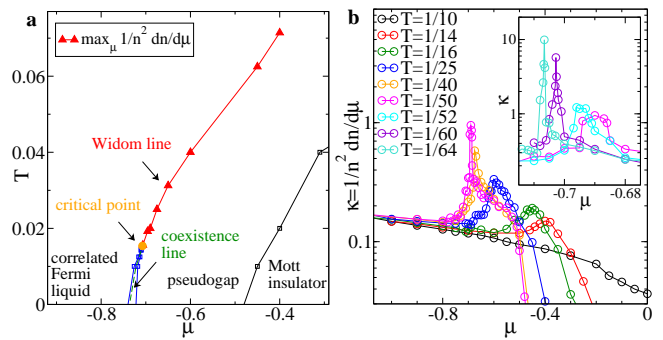}}
\caption{{\bf Phase diagram and Widom line.} (a), temperature $T$ - chemical potential $\mu$ phase diagram of the two-dimensional Hubbard model in the normal-state obtained by cellular dynamical mean-field theory. We take units where $a=t=\hbar=k_B=1$. Here, $U=6.2 > U_{\rm MIT}\approx 5.95$ necessary to create a Mott insulator at half filling ($\mu=0$). The phase transition between two metals, the pseudogap and the correlated Fermi liquid, occurs along the coexistence line (dashed green line) that ends at the critical point $(\mu_{p},T_{p})=(-0.707,1/65)$ (orange circle). The coexistence line is bounded by the spinodals (blue lines with squares). The phase boundaries are obtained from scans at constant $T$ monitoring the behavior of the occupation $n$ versus $\mu$~\cite{sht,sht2}. The onset of the Mott insulator (black line with squares) is defined by a plateau in the occupation at $n=1$. The compressible, metallic, regions have $n<1$. Below $T_{p}$, the particle occupation!
  $n$ changes discontinuously at the spinodal lines, indicating the first-order transition. The green coexistence line and the extrapolations to $T=0 $ are a guide to the eye. The Widom line, i.e. the line where the maxima of different response functions converge, extends the coexistence line above $T_{p}$. We estimate this line by the loci of charge compressibility maxima max$_{\mu} \kappa$, where $\kappa=1/n^2 (dn/d\mu)_T$ (red line with triangles). (b) Semi-logarithmic plot of the charge compressibility $\kappa$ versus chemical potential $\mu$ for several temperatures above $T_{p}$, obtained by numerical derivative of the filling with respect to the chemical potential. The loci of maxima of $\kappa$ are plotted in the $T-\mu$ plane in (a). The value at the maximum increases as $T$ decreases, indicating the divergence of $\kappa$ at the critical point. This is analog to the diverging charge compressibility at the Mott critical point~\cite{sahana} or at the liquid-gas trans!
 ition.}
\label{fig1}
\end{figure}

Recently, we used cellular dynamical mean-field theory on a $2\times 2$ plaquette to demonstrate that a first-order transition inhabits the finite doping part of the normal-state phase diagram of the model~\cite{sht,sht2}.
That first-order transition occurs along a coexistence line between two types of metals and ends at a critical point $(T_{p},\mu_{p})$ (see $T-\mu$ phase diagram in Fig.~\ref{fig1}a).
Here we show that the pseudogap is the low-doping phase whose characteristic temperature $T^*$ can be interpreted as the Widom line.

The Widom line is defined as the line where the maxima of different thermodynamic response functions touch each other asymptotically as one approaches the critical point~\cite{water1}.
The maxima become more pronounced on approaching $T_{p}$, diverging at $T_{p}$.
Fig.~\ref{fig1}b shows the chemical potential dependence of the charge compressibility $\kappa=1/n^2 (dn/d\mu)_T$ for several temperatures above $T_{p}$.
Far above $T_{p}$, $\kappa$ develops a maximum that increases and moves to higher doping with decreasing temperatures, indicating that the charge compressibility diverges at the critical point.
The loci of the charge compressibility maxima, max$|_{\mu} \kappa$, are shown in Fig.~\ref{fig1}a and give an estimate of the Widom line.

Crossing of the Widom line involves drastic changes in the dynamics of the system, as indicated in previous investigations on the phase diagram of fluids~\cite{water1,Simeoni2010}.
Similarly, here we show that the pseudogap, as seen in the single-particle density of states and in the zero frequency spin susceptibility, arises at high temperature from crossing the Widom line that radiates out of the critical point.

%

\begin{figure}
\centering{
\includegraphics[width=0.99\linewidth,clip=]{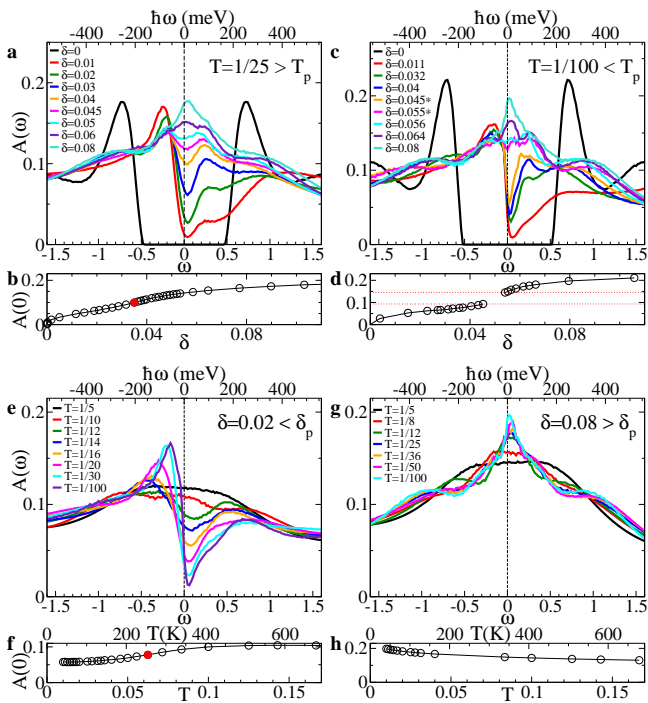}}
\caption{{\bf $T^*$ from density of states.} (a,c,e,g), $A(\omega)$ at $U=6.2$ obtained from maximum-entropy analytic continuation of continuous-time quantum Monte Carlo data. (a,c), $A(\omega)$ for different dopings at fixed temperature, (a) $T =1/25 > T_{p}$  and (c~) $T=1/100 < T_{p}$  respectively. In the latter case, associated with the first-order transition at finite doping there are hysteresis effects in the $T-\mu$ plane, where the transition appears as a coexistence interval between two metallic phases. Orange and magenta lines are $A(\omega)$ obtained in the coexistence region: they have the same chemical potential $\mu=-0.725$, but different filling and different low-energy spectrum, one with a pseudogap and the other one with a rather broad peak at $\omega=0$. (e,g), density of states at fixed doping for various temperatures. (e) is for $\delta=0.02<\delta_{p}$, where a pseudogap opens up with decreasing temperature. (g) is for $\delta=0.08>\delta_{p}$ where a pseudogap is absent and a narrow peak develops as $T$ decreases. (b,d,f,h), density of states at the Fermi level $A(\omega=0)$ as a function of doping or temperature. Here, data are obtained from the extrapolated value of the imaginary part of the local cluster Green's function $-1/\pi $Im$G(\omega_n\rightarrow0)$, which does not require analytical continuation. The inflection point of these curves as a function of $\mu$ or $T$, indicated by a red circle, is our estimate of $T^*$. On the upper horizontal axis we convert into physical units by using $t=0.35eV$.}
\label{fig2}
\end{figure}

{\it Identification of $T^*$ on the basis of the local density of states.}
First we study the development of the pseudogap in the local density of states $A(\omega)$, which can be accessed by tunnelling or photoemission spectroscopy, along paths at constant temperature or at constant doping.
Figs.~\ref{fig2}(a,c) show the evolution of the density of states with doping at a fixed temperature above and below $T_{p}$ respectively.
In the Mott insulating state, at $\delta=0$, the density of states consists of lower and upper Hubbard bands separated by a correlation gap.
Upon hole doping, there is a dramatic transfer of spectral weight from high to low frequency, as a consequence of strong electronic correlations.
The low frequency part of $A(\omega)$ develops a pseudogap, i.e. a depression in spectral weight, between a peak just below the Fermi level and a peak above the Fermi level.
The pseudogap exhibits a two-peak profile that is highly asymmetric~\cite{hauleDOPING}, reflecting the large particle-hole asymmetry observed experimentally~\cite{Davis:2007}.
Upon increasing doping, the particle-hole asymmetry decreases, the spectral weight inside the pseudogap gradually fills in, and the distance between the two peaks slightly decreases.
Finally, at a temperature-dependent doping, the pseudogap disappears and a rather broad peak appears in the density of states which narrows with increased doping.
The change from pseudogap to correlated Fermi liquid behavior occurs either by a first-order transition when $T<T_{p}$, with a discontinuous change in $A(\omega)$, or by a crossover when $T>T_{p}$.
We obtain $T^*(\delta)$ from the inflection point in $A(\omega=0)(\delta)$ at finite doping (see Figs.~\ref{fig2}(b,d)).

Figs.~\ref{fig2}(e,g) show the temperature evolution of the density of states at constant doping.
In Fig.~\ref{fig2}(g) above $\delta_{p}$ (below $\mu_{p}$), there is no pseudogap, only a peak around $\omega=0$ that broadens with increasing temperature~\cite{hauleDOPING,sht2}.
By contrast, in Fig.~\ref{fig2}(e) below $\delta_{p}$ (above $\mu_{p}$), the effect of increasing the temperature is to gradually fill the pseudogap, without decreasing the peak-to-peak distance.
We  identify the disappearance of the pseudogap in the spectrum by the inflection in $A(\omega=0)(T)$ (see Fig.~\ref{fig2}(f) and Supplementary Fig.~S1).
We will discuss later the evolution with doping of $T^*$, but the absence of pseudogap for $\delta>\delta_{p}$, already allows us to conclude that $T^*$ appears as a bridge between the Mott insulator and the first-order transition at finite doping.

%

{\it Pseudogap versus Mott insulator.}
Next we discuss the relationship between the pseudogap and the Mott insulator.
The pseudogap is linked to Mott physics because it opens up only above the critical $U$ for the Mott transition, in the metal near the Mott insulator.
A pseudogap can occur below this threshold due to long-wavelength antiferromagnetic fluctuations, but that is different physics~\cite{Vilk:1997} occurring at a different energy scale.
Here, only short-range spin correlations are involved, as observed experimentally~\cite{Curro:1997} in YBa$_2$Cu$_4$O$_8$ at the pseudogap temperature.
The first-order transition at finite doping, which is the terminus of the pseudogap phase, is linked as well to Mott physics because in the $(U,T,\mu)$ phase diagram, it emerges out of the Mott endpoint at half-filling, and progressively moves away from half-filling with increasing $U$~\cite{sht,sht2}.

\begin{figure}[!h]
\centering{
\includegraphics[width=0.99\linewidth,clip=]{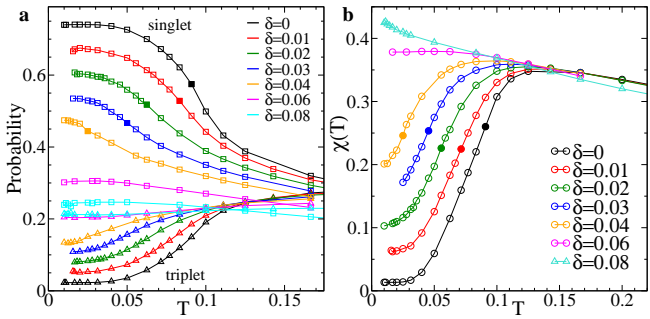}}
\caption{{\bf $T^*$ from spin susceptibility and plaquette eigenstates.} (a), probability of the following plaquette eigenstates as a function of temperature, for several values of doping: the singlet $|N=4,S=0,K=(0,0)\rangle$ and the triplet $|N=4,S=1,K=(\pi,\pi)\rangle$ (squares and triangles respectively), where $N$, $S$, $K$ are the number of electrons, the total spin and the cluster momentum of the plaquette eigenstate.
(b), the zero-frequency spin susceptibility $\chi(T)=\int_{0}^{\beta} \langle S_z(\tau) S_z(0)\rangle d\tau$ as a function of temperature for several values of doping. $S_z$ is the projection of the total spin of the plaquette along the $z$ direction. The inflection point of these curves as a function of $T$, marked by a solid symbol, is our estimate of $T^*$.  Data are for $U=6.2$.}
\label{fig3}
\end{figure}

The pseudogap inherits many features from the parent Mott insulator: in both phases the electrons are bound into short-range singlets because of the superexchange mechanism, reminiscent of the resonating valence bond state~\cite{Anderson:1987}.
Fig.~\ref{fig3}a shows the temperature evolution of the probability obtained from the largest diagonal elements of the reduced density matrix on a $2\times2$ cluster~\cite{hauleCTQMC},
which provide direct access to spin correlations (see also Supplementary Fig.~S2).
Below $\delta_{p}$, upon decreasing temperature, the singlet is gaining weight at the expense of the triplet indicating a reduction of spin fluctuations.
Because of singlet formation, the spin susceptibility $\chi(T)$ drops below a characteristic temperature, as shown in Fig.~\ref{fig3}b and as found in experiments~\cite{Alloul:1989}.
The inflection point in $\chi(T)$ defines a $T^*$ that moves to lower temperatures as the doping increases and approaches $T_{p}$ as $\delta\rightarrow\delta_{p}$.

\begin{figure}
\centering{
\includegraphics[width=0.99\linewidth,clip=]{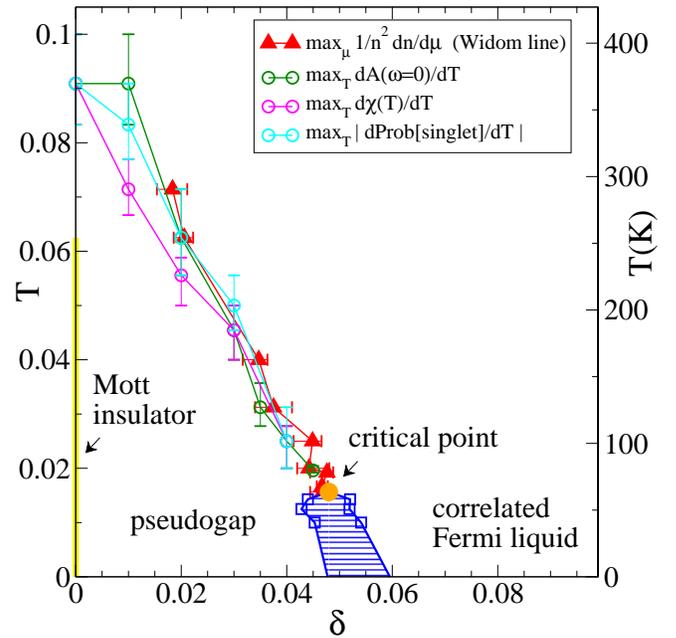}}
\caption{{\bf Pseudogap temperature $T^*$ along the Widom line.} $T-\delta$ phase diagram for $U=6.2$ (see Supplementary Fig.~S3 for the $T-\mu$ phase diagram). The pseudogap temperature scale $T^*$ is computed from the inflection point along paths at constant $\delta$ of $A(\omega=0)$(line with green circles), $\chi(T)$ (magenta circles) and Prob$_{\rm singlet}(T)$ (cyan circles). The $T^*$ line appears as soon as the Mott insulator is doped and joins the critical endpoint $(T_p,\delta_p)$ (orange circle) of a first-order transition away from half-filling, closely following the Widom line (red line with triangles) that emanates from the critical endpoint. The critical point moves to large doping and low temperatures with increasing $U$~\cite{sht,sht2}. The hatched region corresponds to the instability region bounded by the spinodals. Extrapolations to $T=0$ are a guide to the eye. On the right vertical axis we convert into physical units by using $t=0.35eV$.}
\label{fig4}
\end{figure}

Despite the magnetic behavior similar to the Mott insulator phase, the pseudogap phase is a new state of matter.
At $T=0$ it appears to be separated from the Mott insulator by a second-order transition~\cite{sht2}.
With decreasing temperature, neither the value of the density of states at the Fermi level (Fig.~\ref{fig2}(f)), nor the spin susceptibility (Fig.~\ref{fig3}b), extrapolate to zero.
The $T^*$ extrapolated to $\delta=0$ is not related to the opening of the Mott gap (Fig.~\ref{fig4}).
Finally, the peak to peak distance for the pseudogap does not extrapolate, as $\delta\rightarrow 0$, to the Mott gap, just as is observed in experiment~\cite{Hufner:2008}.

%

{\it $T^*$ as the Widom line.}
Now we move to the relationship between the pseudogap and the Widom line emanating from the critical point at finite doping.
Fig.~\ref{fig4} shows in the $T-\delta$ plane the doping evolution of the various $T^*$, identified above as inflection points in $A(\omega=0)$ along constant $T$ or constant $\delta$ paths, and as inflection points in $\chi(T)$ and Prob$_{\rm singlet}(T)$ at constant $\delta$.
The different $T^*$ lines for the pseudogap move closer to each other along the Widom line as we approach the critical endpoint.
The interrelation between $T^*$ and Widom line is our main finding.
Therefore our work shows that the dynamic crossover associated with the buildup of the pseudogap is concomitant with a crossover in the thermodynamic quantities, as observed in supercritical fluids. The organizing principle of these phenomena is the Widom line.
One can thus interpret $T^*$ as the Widom line, or, equivalently, consider the Widom line as a thermodynamic signal of $T^*$.
Our results suggests that all indicators (both thermodynamic and dynamic) of the pseudogap temperature scale $T^*$ should approach each other with increasing doping, joining a critical endpoint of a first-order transition, which thereby appears as the source of anomalous behavior.
In this view, pseudogap and strongly correlated Fermi liquid are separated from each other at low temperature by a first-order transition and are thus two distinct states of matter, just as liquid and gas are two distinct states, or phases.

\noindent{\bf Discussion}\\
Common theories to explain the pseudogap phase include the presence of rotational and/or spatial broken-symmetry phases as an essential ingredient.
By contrast, in our approach the pseudogap is a consequence of large screened Coulomb repulsion that leads to strong singlet correlations in two dimensions reminiscent of resonating valence bond physics~\cite{Anderson:1987}. Competing phases are not necessary to obtain a pseudogap. The pseudogap phase can however be unstable to such phases.
We therefore provide a new and generic mechanism for the pseudogap in doped Mott insulators, according to which the pseudogap state is a new state of matter, whose characteristic temperature $T^*$ corresponds to the Widom line arising above a first-order transition.

The Mott transition is often masked by broken-symmetry states. Similarly, our finite-doping transition is masked by the superconducting phase for instance~\cite{sshtSC}.
Nevertheless the rapid crossover between pseudogap and metallic phases observed above the broken-symmetry states is accessible and can be controlled by the Widom line.
Such rapid change in dynamics is a hallmark of the Widom line~\cite{water1,Simeoni2010} and it is consistent with the strong-coupling nature and the observed phenomenology of the pseudogap in the vicinity of $T^*$ in hole-doped high-temperature superconductors.

From a broader perspective, our work brings the conceptual framework of the Widom line, recently developed in the context of fluids~\cite{water1,Simeoni2010}, to a completely different state of matter, the electronic fluid, suggesting its unexpected generality.
We recall the strong impact that resulted from bringing in the field of electronic properties of solids the well known concepts of smectic and nematic order developed earlier in the field of liquid crystals. It is tempting to argue that the same fate awaits the Widom line.

\noindent{\bf Methods}\\
Our results are based on the cellular dynamical mean-field theory (CDMFT)~\cite{kotliarRMP,maier} solution of the two-dimensional Hubbard model on the square lattice,
\begin{equation}
  H= -\sum_{ij\sigma} t_{ij}c_{i\sigma}^\dagger c_{j\sigma}
  + U \sum_{i}\left(n_{i \uparrow }-\frac{1}{2}\right)\left(n_{i \downarrow }-\frac{1}{2}\right)
  - \mu\sum_{i} n_{i},
\label{eq:HM}
\end{equation}
where $c_{i\sigma}$ and $c^+_{i\sigma}$ operators destroy and create electrons on site $i$ with spin $\sigma$, and $n_{i\sigma}=c^+_{i\sigma}c_{i\sigma}$.
$t$ is the hopping amplitude between nearest neighbors, $U$ the energy cost of double occupation at each site and $\mu$ the chemical potential.
CDMFT isolates a cluster of lattice sites, here a $2\times2$ plaquette, and replaces the missing lattice environment by a bath of non-interacting electrons which is self-consistently determined.
The cluster in a bath problem is solved by a continuous-time Quantum Monte Carlo summation of all diagrams obtained by expanding the partition function in powers of the hybridization between bath and cluster~\cite{millisRMP,hauleCTQMC}.
The size of the plaquette is large enough to be consistent with the experimental observation~\cite{Curro:1997} that at $T^*$, in hole-doped cuprates, the antiferromagnetic correlation length is one or two lattice spacings.

The value of Coulomb interaction $U=6.2t$ is larger than the critical threshold $U_{\rm MIT}\approx 5.95t$ necessary to obtain a Mott insulator at half-filling~\cite{sht,sht2} and is chosen such that the pseudogap critical temperature $T_{p}(U)$ is accessible with our method.
We carry out simulations at constant temperature for several values of $\mu$ and at constant doping $\delta=1-n$ for several temperatures.
Critical slowing down is a widespread and standard signal that the system is approaching a critical threshold~\cite{warnings} and appears in our simulations along the Widom line close to the critical point.
To obtain reliable results, we perform up to $10^7$ Monte Carlo sweeps, averaged over 64 processors, and hundreds of CDMFT iterations.
The typical error on $n$ is of order $10^{-5}$.

{\bf Acknowledgments}
We thank E. Kats and L. Taillefer for discussions. This work was partially supported by FQRNT, by the Tier I Canada Research Chair Program (A.-M.S.T.), and by NSF DMR-0746395 (K.H.). Computational resources were provided by CFI, MELS, the RQCHP, Calcul Qu\'ebec and Compute Canada. A.-M.S.T is grateful to the Harvard Physics Department for support and P.S. for hospitality during the writing of this work. Partial support was also provided by the MIT-Harvard Center for Ultracold Atoms. 

{\bf Author contributions}
G.S. conceived the project and carried the data analysis. P.S. and K.H. wrote the main codes. G.S. and A.-M.S.T. wrote the paper and all authors discussed it. A.-M.S.T. supervised the entire project.

{\bf Competing financial interests}
The authors declare no competing financial interests.

\clearpage

\noindent
{\bf Pseudogap temperature as a Widom line in doped Mott insulators: supplementary information}\\
G. Sordi, P. S\'emon, K. Haule, A.M.-S. Tremblay

\setcounter{figure}{0}
\makeatletter 
\renewcommand{\thefigure}{S\@arabic\c@figure}

\begin{figure}[!h]
\centering
\includegraphics[width=0.95\linewidth]{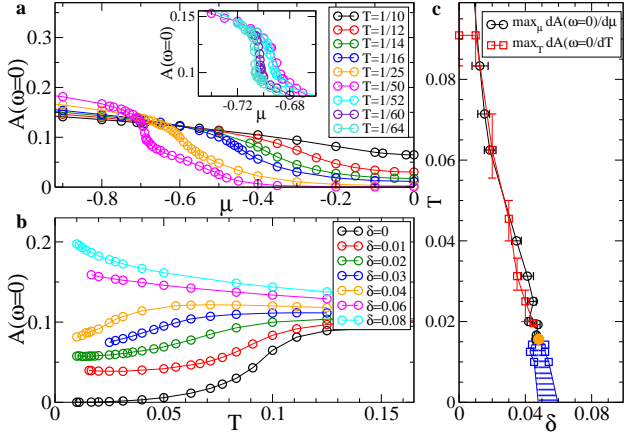}
\caption{(a,b), density of states at the Fermi level $A(\omega=0)$ as a function of (a) chemical potential $\mu$ or (b) temperature $T$. Data are obtained from the extrapolated value of the local cluster Green's function $1/\pi$Im$G(\omega_n\rightarrow0)$. (c~), pseudogap temperature scale $T^*$ in the $T-\delta$ phase diagram. We evaluate $T^*$ by the inflection point in $A(\omega=0)(\mu)$ (line with black circles) and $A(\omega=0)(T)$ (line with red squares). These two estimates of $T^*$ are consistent and terminate at the critical point (orange circle) of the first-order transition. At larger value of $U$~\cite{sht,sht2}, one would obtain a larger value of $\delta_p$, but for large $\delta_p$ the corresponding $T_p$ becomes inaccessible numerically.}
\label{figS1}
\end{figure}

\pagebreak
\begin{figure}
\centering
\includegraphics[width=0.95\linewidth]{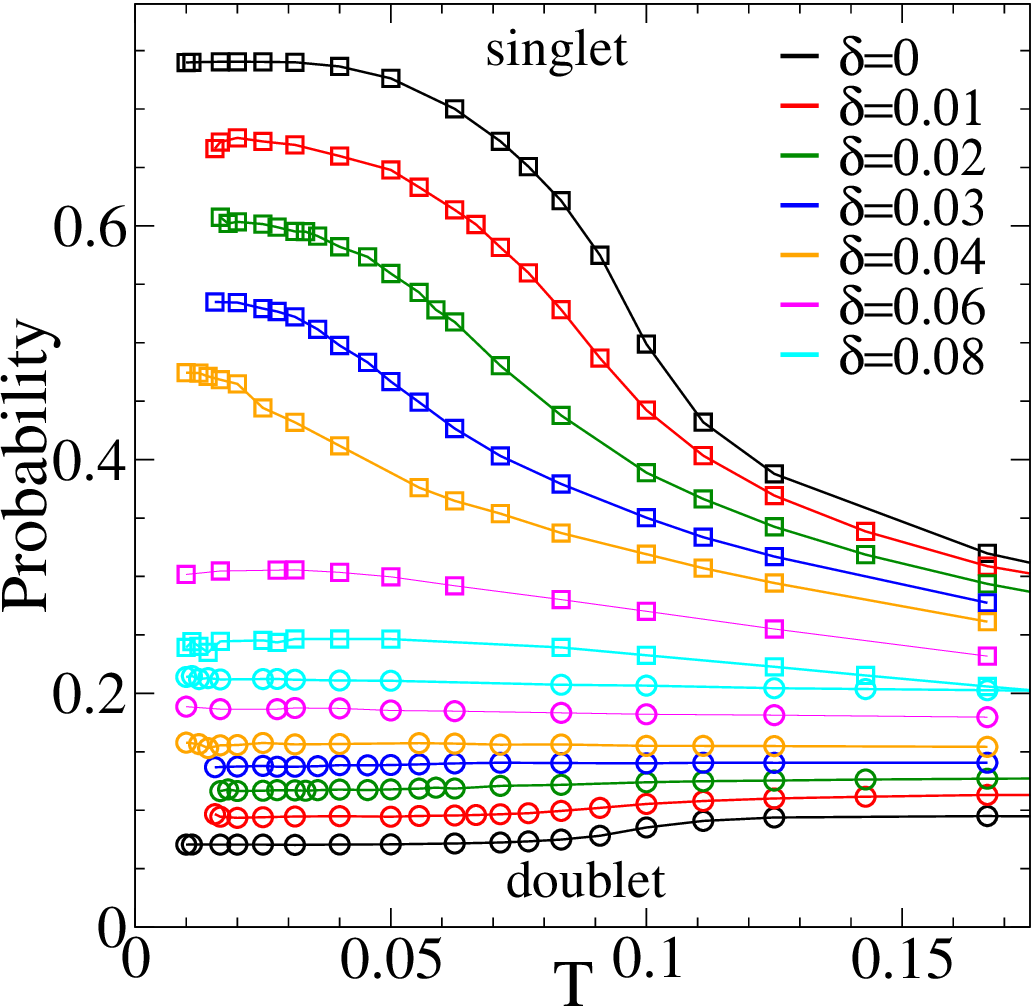}
\caption{Probability of the following $2\times2$ plaquette eigenstates as a function of temperature, for several values of doping: the singlet $|N=4,S=0,K=(0,0)\rangle$ and the doublet $|N=3,S=1/2,K=(\pi,0)\rangle$ (squares and circles respectively), where $N$, $S$, $K$ are the number of electrons, the total spin and the cluster momentum of the plaquette eigenstate. Although the spin doublet is more probable than the triplet over much of the temperature range, its probability is essentially temperature independent.}
\label{figS2}
\end{figure}
\begin{figure}[!h]
\centering
\includegraphics[width=0.95\linewidth]{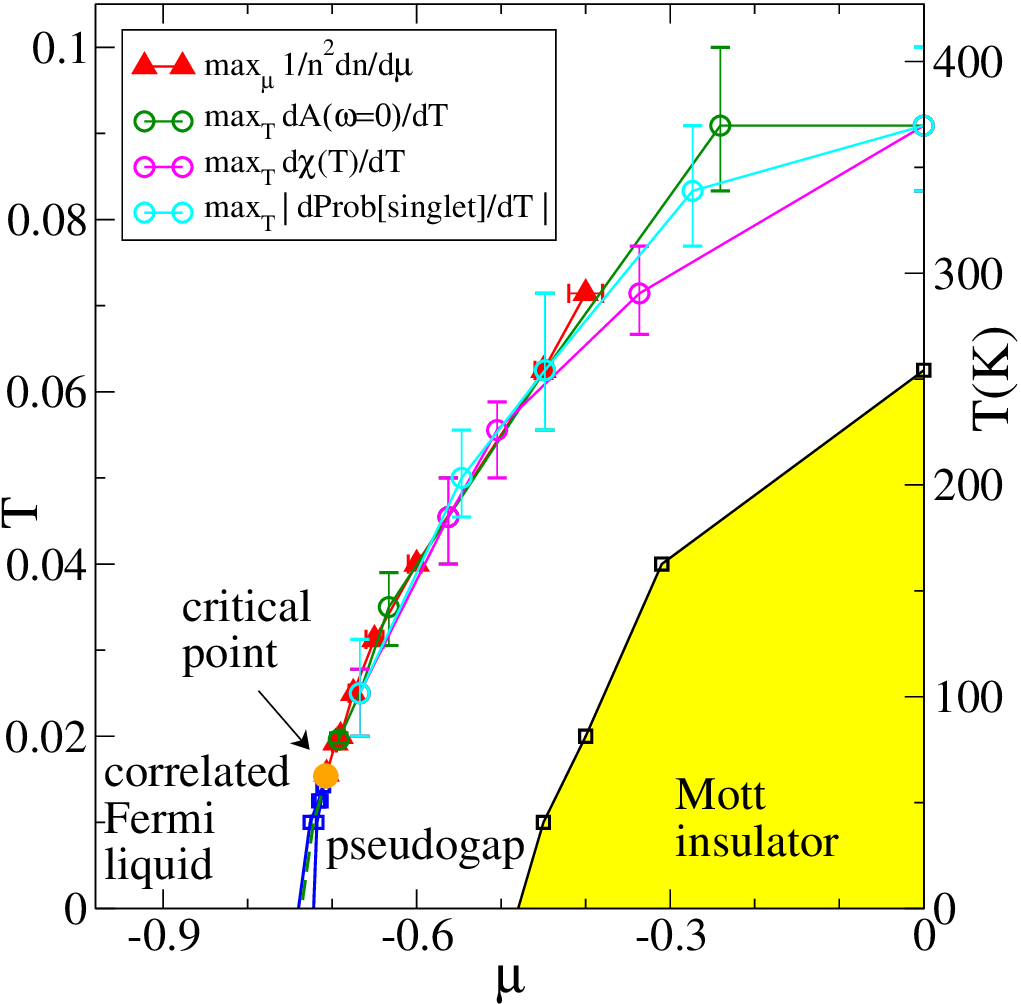}
\caption{$T-\mu$ phase diagram. With respect to Fig.~4, only the x-axis has changed. All indicators of the pseudogap temperature scale $T^*$ follow the Widom line (line with red triangles) and terminate at the critical point (orange circle).}
\label{figS3}
\end{figure}

\end{document}